\documentclass[journal]{IEEEtran}
\usepackage{amsmath,amssymb,amsfonts}
\usepackage{array}
\usepackage[dvips]{graphicx}
\usepackage{enumerate}
\usepackage{subfigure}
\usepackage[svgnames]{xcolor}
\usepackage{float}
\usepackage{algorithm2e}
\usepackage{algorithmic}
\usepackage{multirow}
\usepackage{wasysym}
\usepackage{amsmath,amssymb}
\usepackage{threeparttable}
\usepackage[noadjust]{cite}
\usepackage{graphicx}
\newsavebox{\measurebox}
\usepackage{cite}

\usepackage{array}
\usepackage{threeparttable}

\usepackage{pslatex}
\begin{document}
\title{Proposal of Analog In-Memory Computing with Magnified Tunnel Magnetoresistance Ratio and Universal STT-MRAM Cell}

\author{Hao~Cai,~\IEEEmembership{Member,~IEEE,}
        Yanan~Guo,~\IEEEmembership{Student Member,~IEEE,}
        Bo~Liu,~\IEEEmembership{Member,~IEEE,}
        Mingyang~Zhou,~\IEEEmembership{Student Member,~IEEE,}
        Juntong~Chen,~\IEEEmembership{Student Member,~IEEE,}
        Xinning~Liu,~\IEEEmembership{Member,~IEEE,}
        and~Jun~Yang,~\IEEEmembership{Member,~IEEE}  
\thanks{This work is supported in part by National Key R\&D Program of China under Grant 2018YFB2202800 and National Natural Science Foundation of China under Grant 61904028. (Hao Cai and Yanan Guo contributed equally to this work.)}
\thanks{H. Cai, Y. Guo, B. Liu, M. Zhou, J. Chen, X. Liu and J. Yang are with the National ASIC System Engineering Center, Southeast University, Nanjing 210000, China (e-mail: hao.cai@seu.edu.cn).}
}

\markboth{}%
{Shell \MakeLowercase{\textit{et al.}}: Bare Demo of IEEEtran.cls for IEEE Journals}

\maketitle


\begin{abstract}
In-memory computing (IMC) is an effectual solution for energy-efficient artificial intelligence applications. Analog IMC amortizes the power consumption of multiple sensing amplifiers with analog-to-digital converter (ADC), and simultaneously completes the calculation of multi-line data with high parallelism degree. Based on a universal one-transistor one-magnetic tunnel junction (MTJ) spin transfer torque magnetic RAM (STT-MRAM) cell, this paper demonstrates a novel tunneling magnetoresistance (TMR) ratio magnifying method to realize analog IMC. Previous concerns include low TMR ratio and analog calculation nonlinearity are addressed using device-circuit interaction. Peripheral circuits are minimally modified to enable in-memory matrix-vector multiplication. A current mirror with feedback structure is implemented to enhance analog computing linearity and calculation accuracy. The proposed design maximumly supports 1024 2-bit input and 1-bit weight multiply-and-accumulate (MAC) computations simultaneously. The 2-bit input is represented by the width of the input (IN) pulses, while the 1-bit weight is stored in STT-MRAM and the $\times$7500 magnified TMR (m-TMR) ratio is obtained by latching. The proposal is simulated using 28-nm CMOS process and MTJ compact model. The integral nonlinearity is reduced by 57.6\% compared with the conventional structure. 9.47-25.4 TOPS/W is realized with 2-bit input, 1-bit weight and 4-bit output convolution neural network (CNN). 
\end{abstract}

\begin{IEEEkeywords}STT-MRAM, tunneling magnetoresistance ratio, in-memory computing, analog computing, linearity.
\end{IEEEkeywords}

%
\IEEEpeerreviewmaketitle

\section{Introduction}

\IEEEPARstart{E}{m}erging nonvolatile memories (NVMs) accompany with better computing ways enable great potential for energy-efficient artificial intelligence applications~\cite{top1,top2,top3}. Magnetic random access memory (MRAM) has demonstrated promising developments for in-memory computing (IMC) and near-memory computing (NMC)~\cite{mram-cim1,mram-cim2,mram-cim3}. Compared with other resistive-type NVMs, IMC with MRAM (in-MRAM computing) shows potential with logic compatible supply voltage, relatively small variability issues and high endurance~\cite{mram-cim3}. 

Previous in-MRAM computing realizations mainly rely on intrinsic computing ability of bit-cells~\cite{mram-cim3,SOT1,SOT2,VCMA} and circuit-level modification~\cite{C1,C2,C3,C4}. Circuit-level solutions e.g., Pinatubo, logic-in-memory and computing-in-memory provide Boolean logic computation and support binary neural networks (BNN). MRAM building blocks e.g., sensing amplifier (SA) and reference cells are modified to follow the computation requirement, which could result in deteriorated memory performance. 

Compared with digital IMC, analog IMC shows great merit in terms of high on-chip bandwidth and computation-area efficiency~\cite{SRAM5}. One-transistor one-magnetic tunnel junction (MTJ) 1T-1M bit-cell using spin-transfer-torque (STT) switching mechanism is with foundry supported as the most universal cell structure~\cite{foundry1, foundry2, foundry3}. Unlike other nonvolatile memories, e.g., flash~\cite{flash}, resistive-RAM (RRAM)~\cite{RRAM1,RRAM2} and phase change memory (PCM)~\cite{PCM}, the analogue memory devices behavior of STT-MRAM cannot directly apply to analog computing for energy and throughput constrained applications. The main reason is that 100\%-200\% regular tunnel magnetoresistance (TMR) ratio is overwhelmingly difficult to fulfill the requirement of high dimensional matrix-vector multiplication (MVM). Although TMR was enhanced to 249\%~\cite{TMR}, the discrimination of anti-parallel (AP) and parallel (P) resistance is not sufficient, which can be influenced by the process variations. 

In terms of circuit-level, analog IMC meets obstacles in its development due to: (1) Limited TMR causes mismatch between the impact of external input data and internal stored data on the output. (2) The analog computation behavior is more vulnerable to transistor nonlinear effects than its digital counterpart~\cite{linear}. (3) Multiple bit-lines are required to enable simultaneously for cost amortization. In order to avoid inconsistency operations, processing units must fulfill the calculation requirements of different loads~\cite{linear1}.

In order to bring in-MRAM computing into silicon realization, recent work attempt to make breakthrough through different hierarchical levels. \cite{edlSchottky,Schottky2} integrates MTJ and Schottky diode as a rectified tunnel MR (R-TMR) device. Device fabrication indicates that it is possible to achieve more than 10000\% on/off ratio can be obtained through varying DC offset. \cite{mram-cim3} applies the next-generation spin-orbit torque (SOT) MRAM to realize analog IMC at DNN interface, with Ron equals to 6 MOhms. A STT-MRAM prototype with NMC was demonstrated in~\cite{mram-cim2}. Shift and rotate operations can be realized close to bit-cell array.

In this paper, the proposed a novel device-circuit interacted design approach, as transfer normal TMR (in MTJ device) to magnified TMR (in computation circuits). Analog in-MRAM computing is firstly implemented based on universal 1T-1M cell with STT switching mechanism. To realize an enlarged virtual TMR ratio for analog computing, additional peripheral circuits are implemented, including dynamic latch, current mirror with feedback, and a successive approximation (SAR) analog-to-digital converter (ADC). Nonlinear issues of analog calculation are carefully addressed. The proposal is with minimally modified peripheral circuits realization using 28-nm CMOS, and without modification of bit-cell structure.

The reminder of the paper is organized as follow: Section II explains the basic concept and reviews recent works. Section III presents proposed in-MRAM computing structure. Section IV illustrates simulation results. Finally, Section V compares the performance of this work to prior IMC schemes, and concludes this paper with future remarks.

\section{Preliminaries}
\subsection{Matrix Vector Multiplication}
High-dimensional matrix-vector multiplication (MVM) is a dominant kernel in signal-processing and machine-learning computations~\cite{MVM}. Instead of accessing
raw bits row by row, IMC accesses a computation result over multiple bits, thus amortizing the accessing costs. The structure of the storage array matches the computational form of MVM, so the power consumption can be amortized by implementing MVM with IMC. MVM in-memory realization can split as a combination of multiply accumulate (MAC) computations as: 
\begin{equation}
Y_{OUT} ={\displaystyle \sum\limits _{i=0}^{m-1}} W_{i} \times X_{IN,i} \label{eq1}
\end{equation}
where $W$ is the weights stored in memory array, $X_{IN}$ is the activations from external input, and $Y_{OUT}$ is the calculation results. There are $m$ weights and $m$ activations involved in the MAC calculation. 

MVM needs to map the parallel input data to each row, map the parallel computation to the bit cells where the data is stored, and then reduce the output data by adding up. The mapping of weights in the storage array can be represented by:
\begin{equation}
W=\begin{pmatrix}
W_{0,n-1} & W_{0,n-2} & \cdots  & W_{0,0}\\
W_{1,n-1} & W_{1,n-2} & \cdots  & W_{1,0}\\
\vdots  & \vdots  & \ddots  & \vdots \\
W_{m-1,n-1} & W_{m-1,n-2} & \cdots  & W_{m-1,0}
\end{pmatrix} \label{eq2} 
\end{equation}
where $n$ means each weight is quantified to $n$ bits.

\subsection{Universal 1T-1M bit-cell}

\begin{figure}[ht]
    \centering{
    \includegraphics[width=0.4\textwidth]{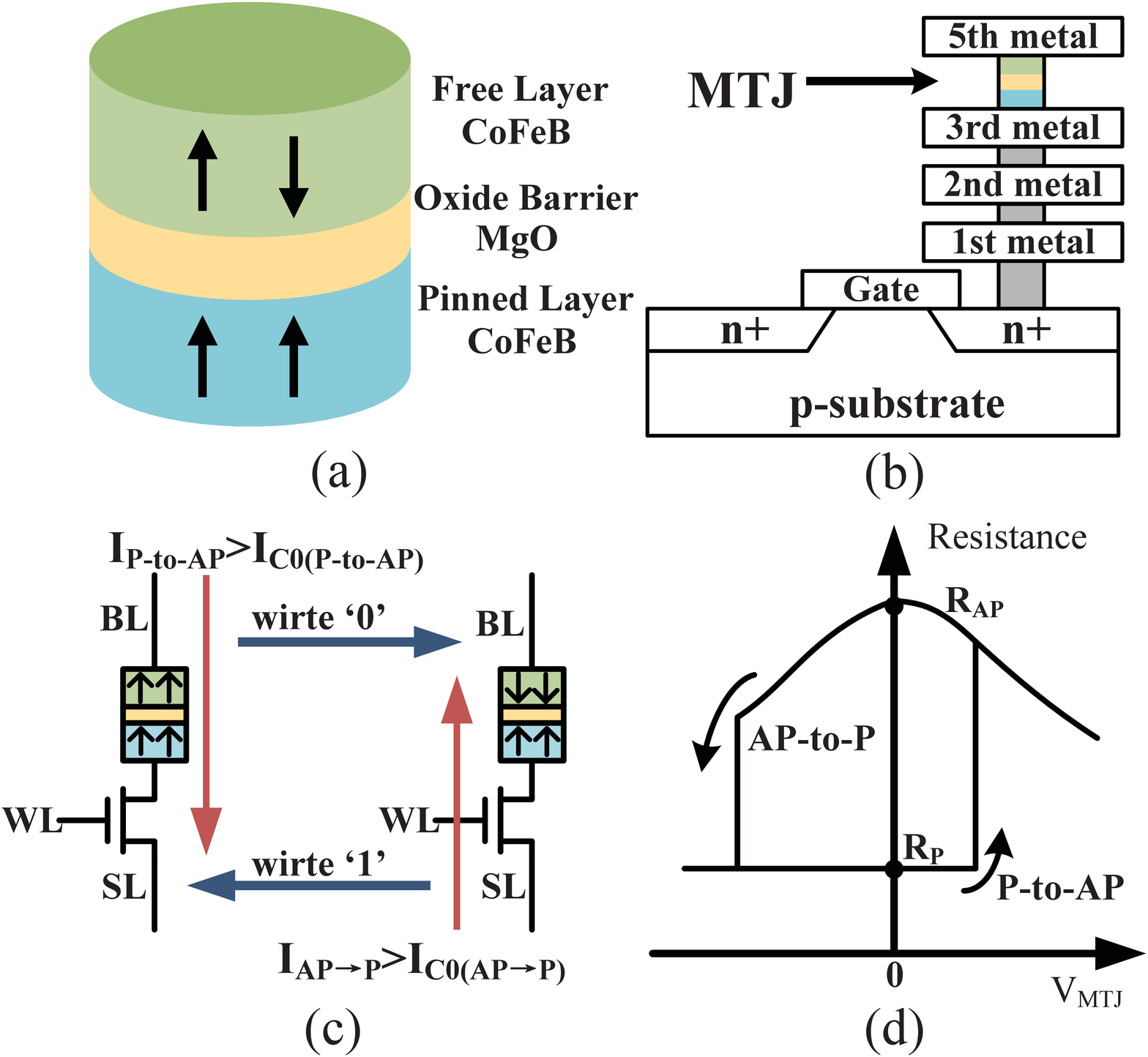}
    \caption{(a) Sandwich structure of PMA-MTJ. (b) Cross-sectional view of 1T-1M bit-cell. (c) Universal 1T-1M bit-cell structure. (d) Resistance loop of MTJ measured by DC voltage sweep.}
    \label{fig:1}}
\vskip -2pt
\end{figure}

Foundry-support 1T-1M bit-cell structure shows high-density and access energy efficiency in MRAM array~\cite{foundry1, foundry2, foundry3}. Fig.~\ref{fig:1}(a) illustrates the basic storage element of perpendicular magnetic anisotropy (PMA) MTJ. Compared with in-plane magnetic anisotropy (IMA),  PMA MTJ fulfills the thermal stability requirement, but also has no restriction of cell aspect ratio, which shows the scaling advantages of high-density integration~\cite{PMA1,PMA2}. MTJ consists of two ferromagnetic electrodes (CoFeB) with a tunnel barrier layer (MgO). The top magnet is the storage layer (referred as the free layer) and the bottom magnet is the reference (referred as the pinned layer). Nonvolatile data writing is performed by injecting a spin-polarized current from one of the ferromagnetic electrodes to change the magnetic orientation of the free layer. The effective resistance of MTJ is low ($R_{P}$) when two ferromagnetic electrodes are spin aligned, and high ($R_{AP}$) when the magnetic direction of two layers is in the anti-parallel state. The resistance difference is represented as TMR ratio~\cite{TMR,MTJ1,MTJ2}:
\begin{equation}
TMR=\frac{R_{AP} -R_{P}}{R_{P}}.
\label{eq3} 
\end{equation}

Fig.~\ref{fig:1} demonstrates the cross-sectional view, bit-cell structure and I-V characteristics of 1T-1M bit-cell. According to spin transfer torque mechanism, a bidirectional current can change the MTJ between states when it is higher than critical current $I_{C0}$. During the write operation, the current flowing from bit line (BL) to source line (SL) can write the data stored in MTJ as `0', while the current with opposite direction can write `1'. Fig.~\ref{fig:1}(d) shows the resistance state at different voltages and the transition of MTJ between P state and AP state.

\subsection{Digital and analog realization of IMC}

\begin{figure}[ht]
    \centering{
    \includegraphics[width=0.48\textwidth]{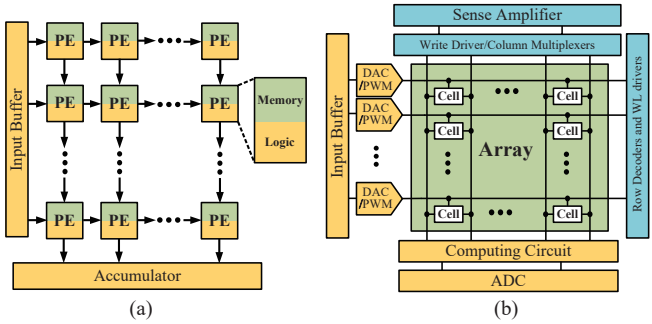}
    \caption{Conventional IMC structure (blue blocks are read-write modules; green blocks are memory modules; orange blocks are digital or analog computing modules): (a) Digital IMC. (b) Analog IMC.}
    \label{fig:2}}
\vskip -2pt
\end{figure}

IMC is an effective way to achieve energy-efficiency emphasized non-von Neumann architecture. The principle operation of IMC is the MAC step. The few-updated weight data is stored within the memory array, and the data to be proceeded is the input from the outside. Completing part or all of the calculation in memory can reduce the power consumption caused by data transfer in the calculation. According to computation signal type and implementation approach, IMC can be classified as digital~\cite{SRAM1,SRAM2} and analog~\cite{SRAM3,SRAM4,DRAM,flash,RRAM1,RRAM2} computing. The framework of digital and analog IMC are illustrated in Fig.~\ref{fig:2}. 

\begin{table*}[!t]
\centering
\caption{Comparison of recent IMC (full chip) for neural network}
\setlength{\tabcolsep}{3pt}
\begin{threeparttable}
\begin{tabular}{|m{42pt}<{\centering}|m{40pt}<{\centering}|m{43pt}<{\centering}|m{42pt}<{\centering}|m{39pt}<{\centering}|m{39pt}<{\centering}|m{39pt}<{\centering}|m{42pt}<{\centering}|m{39pt}<{\centering}|m{39pt}<{\centering}|m{39pt}<{\centering}|}
\hline
& JSSC'21  & ISSCC'21  & JSSC'19   & ISSCC'20   & ISSCC'21  & ISCAS'21  & Nat.Com'20 & ISSCC'20  & ISSCC'21  & ISCAS'19\tnote{1} \\
&  \cite{SRAM1} & \cite{SRAM2} & \cite{SRAM3} & \cite{SRAM4} & \cite{DRAM}  & \cite{flash} & \cite{PCM}& \cite{RRAM1} & \cite{RRAM2} & \cite{STT-MRAM}\\\hline
Domain & Digital & Digital & Analog & Analog & Analog & Analog & Analog & Analog & Analog & Analog\\\hline
Type &SRAM  &  SRAM&SRAM  &  SRAM   &   DRAM  &   Flash  & PCM &   RRAM & RRAM & STT-MRAM  \\\hline
Process & 65nm & 22nm &65nm &7nmFinFET &65nm & 180nm   & 90nm & 130nm   &   22nm  &   22nm  \\\hline
Bit-cell & 6T+XNOR +MUX+FA & 6T+NOR &8T+1C  &   8T   &  1T1C  &  1T1FTG  & 1PCM &  2T2R & 1T1R & 1T-1M  \\\hline
Capacity  &16KB$\times$8 &64Kb & 36Kb$\times$64 & 4Kb & 16Kb & 575Mb & 1Mb & 158.8Kb & 4Mb & 36Kb \\\hline
Macro Area ($\displaystyle \mathrm{mm^2}$) &1 &0.202 & 12.6 & 0.0032 & 0.57 & 17.3 & N/A & 21.82 & 6 & N/A \\\hline
Throughput (GOPS) & 6.1-567 &3300 (4b/4b) 24.7 (8b/8b)& 18876 & 372.4 & 4.71 & 29000 frame/sec\tnote{2} & N/A & N/A & 35.59-417.96 & N/A \\\hline
Weight Bit & 1-16 & 4/8/12/16 & 1 & 4 & 8(signed) & 4 & 4 & 2-3(signed) & 1/4/8 & 5 \\\hline
Input Bit &1-16& 1-8& 1 & 4 & 8 & 4 & 8 & 1 & 2/4/8 & 4 \\\hline
Output Bit & 1-16& 16 (4b/4b) 24 (8b/8b)& 1 & 4 & 8 & 8 & 8 & 1-8 & 4/10/14 & 4 \\\hline
Computation & MAC,DNN & MAC & Binary CNN & MAC & MAC,CNN & MAC,DNN & MAC,DNN & MAC & MAC & MAC,DNN \\\hline
Energy Effi. (TOPS/W) & 2.06-117.3 & 89 (4b/4b) 24.7 (8b/8b) & 866 & 351\tnote{3} & 4.76 & 37\tnote{4} & 11.9 & 78.4 & 11.91-195.7 & 9.7\tnote{5} \\\hline
\end{tabular}
 \begin{tablenotes}
        \footnotesize
        \item[1] Simulation results only without measurement data.~~~~~~~~~~~~~~~~~~~~~~~~~~~~~~~ $^2$ The computing throughput for VGG7.
        \item[3] Each 4b $\times$ 4b is considered as 2 operations.~~~~~~~~~~~~~~~~~~~~~~~~~~~~~~~~~~~~~~~$^4$ MAC only, not including other peripheral work.
        \item[5] It is estimated from the data in the paper.
      \end{tablenotes}
 \end{threeparttable}
\label{Table1}
\vskip -10pt
\end{table*}

Digital IMC merges embedded memories and Boolean logic block to form processing element (PE) unit. The acceleration array is composed of multiple PE units. Input buffer and accumulator are responsible for data input and output, respectively. 

Analog IMC retains the structure of the memory array and read-write function. External activation signals are input to each word-line (WL) through digital-to-analog converter (DAC) or pulse width modulation (PWM), and multiple rows are accessed at once. The DAC converts a digital quantity of input into a WL voltage value or the PWM adjusts the time for the high level, thus controlling the WL of each bit-cell and modulating the bit-cell current. Within the bit-cell, the weight data stored in the memory and the external input data are completed the AND operation to achieve the multiplication operation. According to Kirchhoff current law, the current flowing through each bit-cell is summarized on the bit line (BL). The result of the accumulation operation is expressed as the BL voltage drop or the voltage value on the output capacitor. Finally, an ADC converts the voltage amplitude into the digital output of memory-computation integrated block. Depending on the trade-off between precision and energy consumption, 4-8 bit ADCs are usually used in analog computing~\cite{PCM,RRAM1}.

Table~\ref{Table1} lists a literature study of recent IMC, including various types of memory and calculation methods. Compared with analog IMC, the digital way has the advantage of computational flexibility and accuracy. But its approach of combining logical units with bit-cells requires a lot of area overhead. The advantage of analog IMC is with high parallelism degree, which can bring enhanced throughput and energy efficiency than the digital way. Nevertheless, the design challenge of analog IMC is the design trade-off among accuracy, additional layout area and power consumption of analog components~\cite{SRAM5}. 

The main advantages of using static RAM (SRAM) for analog IMC are mature technology and fast computing speed. The binary logic stored in SRAM can be easily distinguished, allowing for high margins and computational linearity when multiple lines are turned on. The enhanced 8T SRMA bit-cell can effectively reduce the interference of computing to the storage data\cite{SRAM3,SRAM4}. \cite{SRAM3} uses the charge sharing of capacitors to realize the addition operation, reducing the power consumption caused by the BL current and improving the energy efficiency. In \cite{DRAM}, the columns of DRAM are configured as charge-sharing cells, and the dot product operation is accomplished by non-destructive weight reading, saving peripheral circuit area and power consumption.

NVMs, e.g., Flash, PCM and RRAM show potential for analog computing because of high resistance ratio. \cite{RRAM1} stores multi-bit signed numbers in a 2T-2M bit-cell, and uses ADC with adjustable precision to meet different computing requirements. In \cite{RRAM2}, the multi-bit input is split and computed. Under the premise of ensuring accuracy, this method implements MAC with multi-bit I/O unit. 

MRAM cannot be directly used for analog IMC due to limited TMR. \cite{STT-MRAM} proposes a new analog computing structure with operational amplifier integrator circuit, which uses the difference of the current flowing through $R_{P}$ and $R_{AP}$ to complete the calculation. This method alleviates the problem of limited TMR, but it has a high requirement on the performance of operational amplifier.

\section{Proposal of Analog In-MRAM Computing}

This section aims to solve major analog IMC design challenges: limited device TMR ratio, circuit nonlinearity and multi-bit computation configuration. The proposal refers to minimally modified memory array circuits. That is, universal bit-cell is used and the majority of MRAM peripheral circuits is retained. The dual-mode bit-cell is configured to realize magnified TMR ratio (m-TMR) for analog computing. Integrating current mirror with feedback can enhance the linearity of the analog calculation.

\begin{figure}[ht]
    \centering{
    \includegraphics[width=0.48\textwidth]{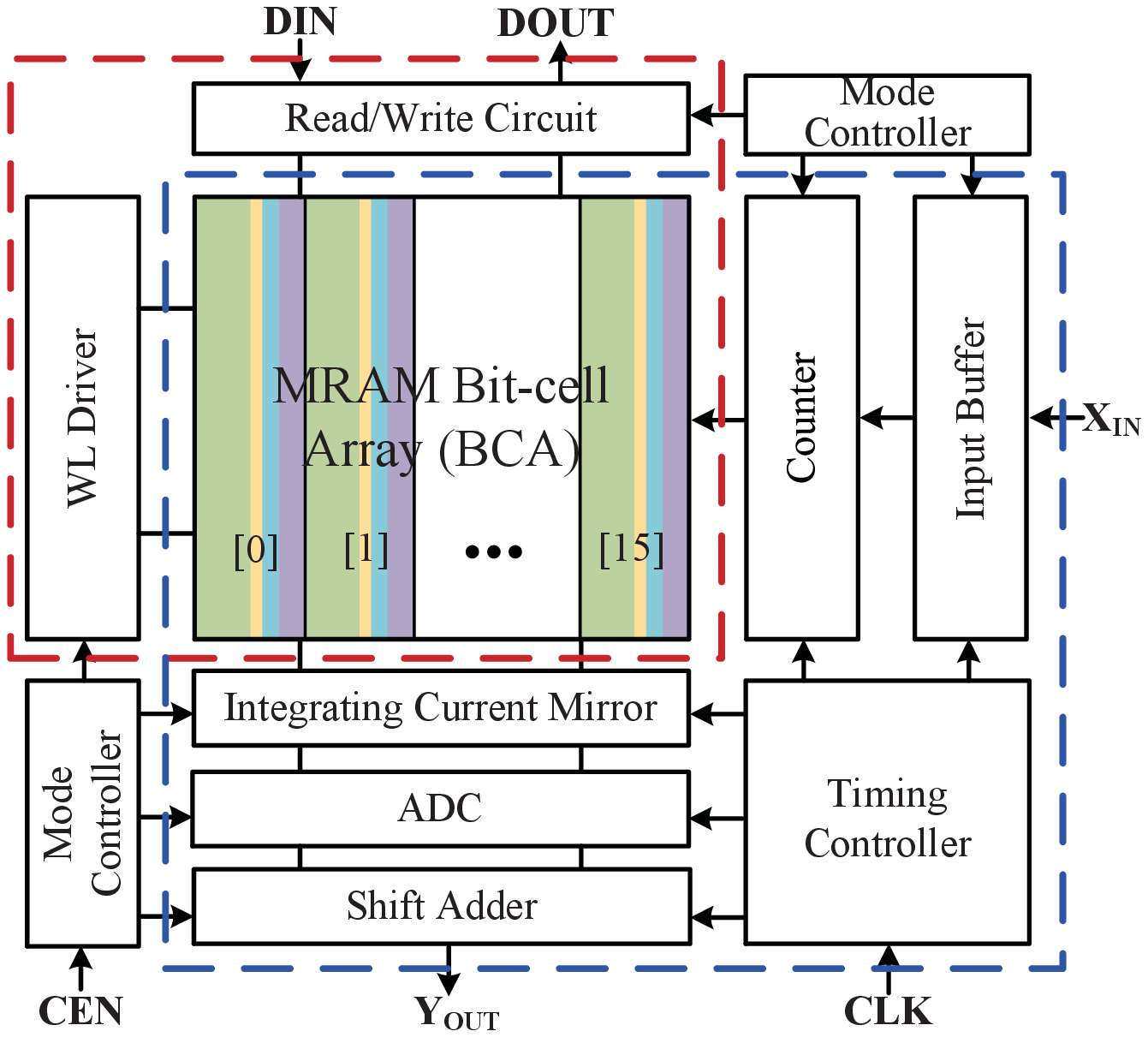}
    \caption{Floorplan of proposed in-MRAM computing. Modified (additional) peripheral blocks mainly include: dynamic latch, current mirror with feedback and SAR-ADC. }
    \label{fig:3}}
\vskip -2pt
\end{figure}

\subsection{The floorplan of dual-mode MRAM}

Fig.~\ref{fig:3} illustrates the block diagram of in-MRAM computing. The operation includes normal storage and IMC mode. A mode controller selects the dual-modes according to $CEN$ (computing enable) signal. In the storage mode, the MRAM bit-cell array (BCA) executes the read-write operation as standard memory. Alternatively, computing mode is executed with analog MVM, which consists of 16 MAC operations in equation~(\ref{eq1}). As shown in Fig.~\ref{fig:3}, the MRAM BCA is divided into 16 local BCAs, each corresponding to 16 MAC operations. $W$ is stored in the local BCA as shown in equation~(\ref{eq2}). And $X_{IN}$ is the activations from external input, which is converted into pulse signal of corresponding width after passing the counter. The multiplication is completed in the BCA, the accumulation is completed in the integrating current mirror. Finally, $Y_{OUT}$ is calculated using a SAR-ADC and shift adders. 

\subsection{The dual-mode bit-cell}

\begin{figure}[ht]
    \centering{
    \includegraphics[width=0.37\textwidth]{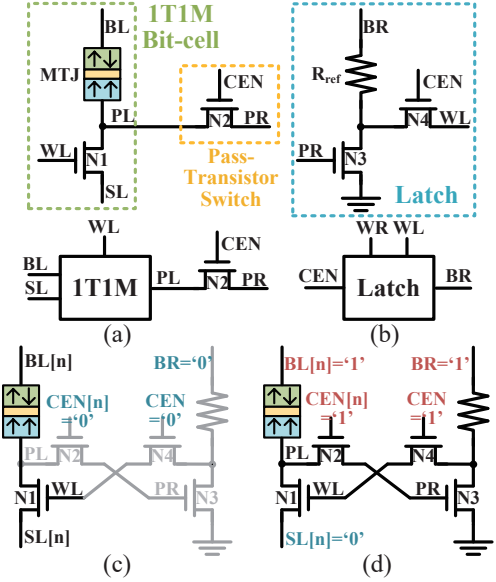}
    \caption{(a) Structure of the 1T-1M bit-cell with a pass-transistor (PT) switch for in-MRAM computing. (b) Structure of the latch unit. (c) Normal storage mode. (d) In-MRAM computing mode.}
    \label{fig:4}}
\vskip -2pt
\end{figure}

As shown in Fig.~\ref{fig:4}(a), 1T-1M bit-cell is applied to normal storage mode. The pass-transistor (PT) switch N2 is allocated for in-MRAM computing mode. Compared with universal 1T-1M cell, a PT interface $PL$ is included to initiate in-MRAM computing. The 1T-1M bit-cell structure with PT switch is regulated with $BL$, $WL$, $SL$, $PR$ and $CEN$ signals. The latch unit is consisted of two transistors and a reference resistance ($R_{ref}$), as shown in Fig.~\ref{fig:4}(b). The resistance value of $R_{ref}$ is setup between $R_{P}$ and $R_{AP}$ of MTJ. This specific value is determined according to the simulation results of power consumption and latch yield, which will be elaborated in Section IV. In the local BCA, latch unit and 1T-1M cell with PT switch are connected by $WL$ and $PR$. The $CEN$ of two cells control the mode of local BCA. 

Fig.~\ref{fig:4}(c) shows the working principle of the storage mode. The $CEN$s of two units are both low levels. Transistor N2 and N4 are cut-off, the proposed 1T-1M bit-cell is identical to conventional 1T-1M. It can complete read and write operations. In storage mode, $W$ is stored in the local BCA and is the preparation for the in-MRAM computing. Compared to other storage such as SRAM, non-volatile MRAM has no static power consumption while holding data.

\begin{figure*}[ht]
    \centering{
    \includegraphics[width=0.95\textwidth]{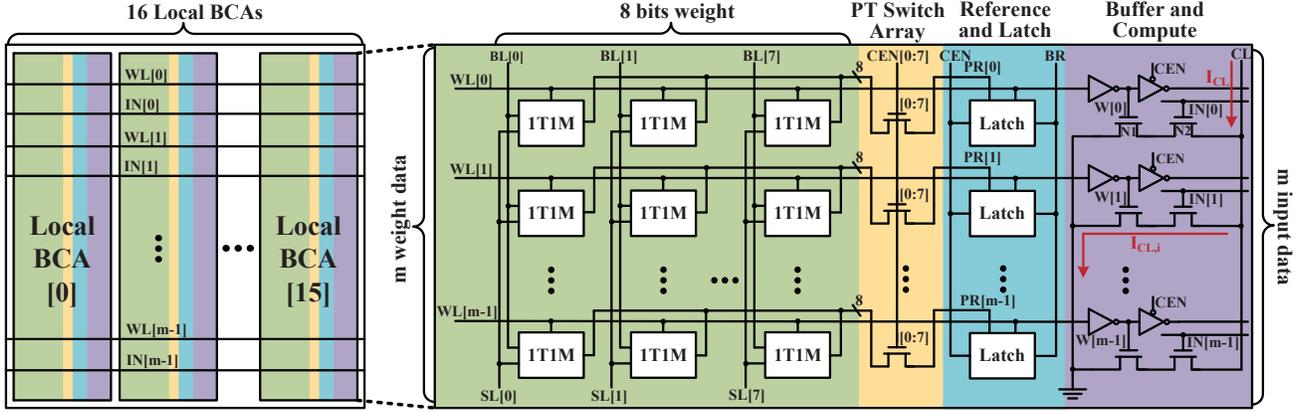}
    \caption{The block diagram of local BCA. It mainly includes memory array using universial 1T-1M bit-cells, as well as additional PT switch array, reference-latch column and buffer-compute column in-MRAM computing.}
    \label{fig:5}}
\vskip -2pt
\end{figure*}

When $CEN[n]$, $CEN$, $BL$, $BR$ are high and $SL$ is low, two cross connected units are in the computing mode. As shown in Fig.~\ref{fig:4}(d), the $CENs$ of two units are both high, and two units are connected to form a latch structure. According to the resistance value of MTJ, high and low levels are generated on $WL$ and $PR$ to complete the latching operation. The magnetic orientation of the pinned layer is away from the transistor N1. When the resistance of MTJ is higher than $R_{ref}$, $PR$ is with low level and $WL$ is high. The MTJ is in AP state, and the magnetic orientation of the free layer is the same as the current. The high voltage difference between the two ends of the MTJ does not affect the data stored in the MTJ.

\subsection{Local bit-cell array}

\begin{figure}[ht]
    \centering{
    \includegraphics[width=0.45\textwidth]{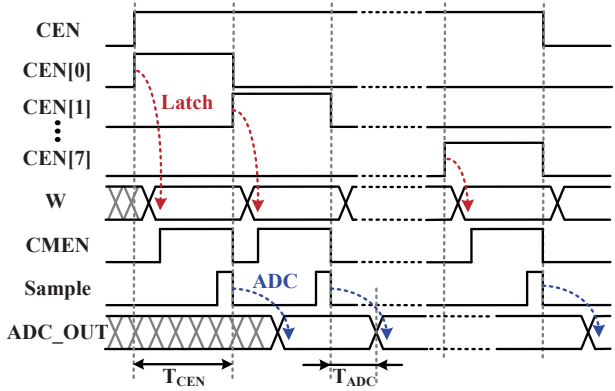}
    \caption{Timing diagram of local BCA.}
    \label{fig:10}}
\vskip -2pt
\end{figure}

Fig.~\ref{fig:5} shows the block diagram of local BCA, which consists of the proposed 1T-1M dual-mode cell, PT switch array, reference-latch column and buffer-compute column. The $WL$ connect all columns in the local BCA, and the $WL$ of 16 local BCAs are isolated by the buffer of the buffer-compute column. The PT switch array controls connection between each bit-cell and the latch unit. In the computing mode, each column of the 1T-1M bit-cell array is connected with the reference-latch column in turn to generate high and low levels on each $WL$ using the latch structure, as shown in Fig.~\ref{fig:10}. The buffer-compute column can prevent the local BCA from interfering with each other in the computing mode. Meanwhile, the buffer-compute column can amplify the weight signal on the WL to node W using the first stage inverter of the buffer. MTJ resistance is converted into the gate voltage of transistor N1 by the latch structure. The TMR is ultimately amplified to the m-TMR: 
\begin{equation}
m\text{-}TMR=\frac{R_{off} -R_{on}}{R_{on}}
\label{eq4}
\end{equation}
where $R_{on}$/$R_{off}$ is the on/off resistance of transistor. 

N1 and N2 are connected in series to complete the multiplication of $W$ and $IN$. $IN$ is the input pulse generated by the activations from external $X_{IN}$. The voltage is provided in the computing line ($CL$) of buffer-compute column, and different current $I_{CL,i}$ is generated in each row, which are calculated through $CL$ and subsequent modules. According to Kirchhoff current law, the $CL$ current $I_{CL}$ is given by:
\begin{equation}
I_{CL} ={ \sum\limits _{i=0}^{m-1} I_{CL,i}}
\label{eq5}
\end{equation}
where $I_{CL,i}=I_a$ when W[i] and IN[i] are high, otherwise $I_{CL,i}=0$. The computing port $CL$ and the read-write port $BL$ are decoupled, which can improve the computation stability and the amount of data that can be simultaneously accessed~\cite{decouple}.

\subsection{The associated current mirror and ADC}

A local BCA corresponds to an integrating current mirror and a SAR-ADC. Fig.~\ref{fig:10} shows the operation waveform of these modules. $CMEN$ the enabling signal of integrating current mirror. $Sample$ is the sampling signal of SAR-ADC, which overlaps with the $CMEN$ to improve the speed of in-MRAM computing. At the end of sampling, SAR-ADC converts the sampling voltage, while the array continues to calculate the next column. The operation mode of pipeline can greatly shorten the calculation delay. The calculation delay of one column is given by:
\begin{equation}
T ={ T_{CEN} + T_{ADC}}
\label{eq6}
\end{equation}
where $T_{CEN}$ indicates the latch and compute time, and $T_{ADC}$ is the conversion time of ADC. The more columns that are calculated, the less the average delay per column. The average delay of n columns is given by:
\begin{equation}
{T_{AVG}} ={ T_{CEN} + \frac{1}{n}\times T_{ADC}}.
\label{eq7}
\end{equation}

\subsection{Operations of Input and Output}
Table~\ref{Table2} lists the multiplication of 2-bit input and 1-bit weight. $X_{IN}$ is the input via buffer, then the counter generates a set of pulses whose width $T_{IN}$ corresponds to the value of the activation signal. This set of pulse is connected to the $IN$ of the buffer-compute column (see Fig.~\ref{fig:4}), and determines the calculated current conduction time for each row of the buffer-compute column. $T_{CP}$ is the period of the $CP$ pulse. $V_a$ represents the output voltage corresponding to the output result of '1'.

\begin{table}[ht]
\centering
\caption{Multiplication of 2-bit input and 1-bit weight}
\setlength{\tabcolsep}{3pt}
\begin{tabular}{|p{30pt}<{\centering}|p{30pt}<{\centering}|p{30pt}<{\centering}|p{30pt}<{\centering}|p{30pt}<{\centering}|}
\hline
MTJ  & $W$ & $X_{IN}$  & $T_{IN}$ & $V_{OUT}$ \\\hline
\multirow{4}{*}{P}  &  \multirow{4}{*}{1}  &   00  &   0  &   0  \\\cline{3-5}
  &   &   01  &   $1 \times {T_{CP}}$  &   $1 \times {V_a}$  \\\cline{3-5}
  &   &   10  &   $2 \times {T_{CP}}$  &   $2 \times {V_a}$  \\\cline{3-5}
  &   &   11  &   $3 \times {T_{CP}}$  &   $3 \times {V_a}$   \\\hline
AP   &   0  &   X  &  X  &   0  \\\hline
\end{tabular}
\label{Table2}
\end{table}

\begin{figure}[ht]
    \centering{
    \includegraphics[width=0.45\textwidth]{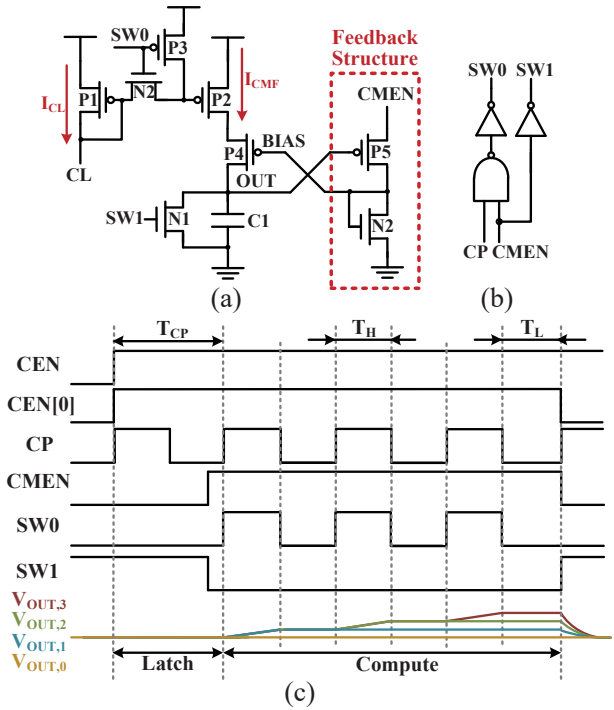}
    \caption{(a) Circuit of integrating current mirror with feedback (CMF) structure. (b) The timing control logic circuit. (c) Waveform of computation control signals. $V_{out,1/2/3/4}$ is the output waveform of different $X_{IN}$.}
    \label{fig:6}}
\vskip -2pt
\end{figure}

Fig.\ref{fig:6}(a) is the circuit of integrating current mirror with feedback (CMF) structure. When reading multiple row data, the current mirror can be used to keep the read current constant, thus enhancing the calculated linearity~\cite{linear}. The current $I_{CL}$ is mirrored as $I_{CMF}$ by the CMF structure, and the voltage of CL does not change with current charge. $I_{CMF}$ is given by:
\begin{equation}
{I_{CMF}} ={ \gamma \times I_{CL}}
\label{eq8}
\end{equation}
\begin{equation}
{\gamma} ={ \frac{W_{P2}/L_{P2}}{W_{P1}/L_{P1}}}.
\label{eq9}
\end{equation}
Then $I_{CMF}$ is charged to the capacitor and converted into output voltage ($V_{OUT}$). The charge current decreases with the increase of $V_{OUT}$, which reduces the linearity of the analog calculation and affects the calculation accuracy. Proposed feedback structure of the CMF structure can reduce the bias voltage ($V_{BIAS}$) when $V_{OUT}$ goes up. The charge current of the capacitor C1 is stabilized by feedback. 

The timing control logic and the operation waveform of CMF structure is demonstrated in Fig.~\ref{fig:6}(b) and \ref{fig:6}(c). $SW0$ controls the charging time ($T_H$) of each cycle, and the width of $IN$ pulse only controls whether to charge. The $SW0$ of CMF structure can provide more accurate unified control of charging time than the pulse of $IN$. The three pulses of $SW0$ correspond to the 2-bit $X_{IN}$. $V_{OUT,i}$ is the waveform with the corresponding $X_{IN}$ value of $i$ for the MAC of 1 weight ($W=1$) and 1 activation. By adjusting the number and width of SW0 pulses, CMF structure can satisfy $X_{IN}$ of different bits. For the MAC of m weights and m activations, $V_{OUT}$ can be represented as:
\begin{equation}
{V_{OUT}} ={ \sum\limits _{i=0}^{m-1} W_{i} \times X_{IN,i} \times V_a}
\label{eq10}
\end{equation}
\begin{equation}
{V_a} ={ \frac{1}{C} \times \gamma \times I_a \times T_H}
\label{eq11}
\end{equation}
where $C$ is the capacitance of capacitor C1.

\subsection{SAR-ADC}
\begin{figure}[ht]
    \centering{
    \includegraphics[width=0.48\textwidth]{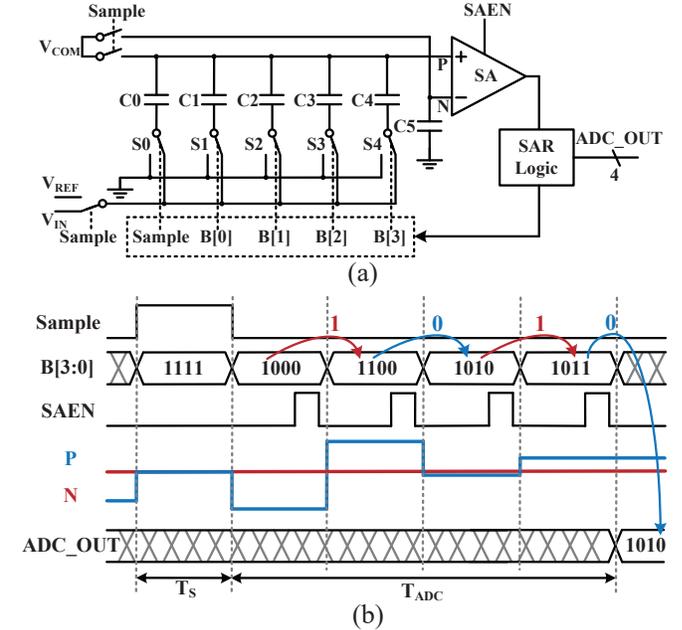}
    \caption{(a) Schematic of 4-bit SAR-ADC. (b) Operation waveform of SAR-ADC, when ADC-OUT is `1010'.}
    \label{fig:11}}
\vskip -2pt
\end{figure}

SAR-ADC is commonly used in analog IMC because of its low cost in layout area and power consumption for multi-bits conversion. In this work, we use a 4-bit SAR-ADC as the an important peripheral circuit of the analog in-MRAM computing. Fig.~\ref{fig:11}(a) presents a schematic of the 4-bit SAR-ADC comprising five weighted capacitors (C4-C0), five switches (S4-S0), SA and SAR logic for binary search algorithm. The capacitance ratio of C4, C3, C2, C1, and C0 is 8:4:2:1:1. The top plates of the five capacitors are shorted to the node $P$ of the SA, and the total parasitic load at node $P$ is 16 times that of C0.

Fig.~\ref{fig:11}(b) illustrates an example of the readout operation of 4-bit SAR-ADC. At the beginning of the readout operation, the voltages at node $P$ ($V_P$) and node $N$ ($V_N$) are initially set to a common voltage $V_{COM}$ ($V_P = V_N = V_{COM}$). After setting the input node to the analog input voltage $V_{IN}$, all five switches (S4-S0) are switched from $V_{IN}$ to ground to decrease $V_P$ by $(16/16) \times V_{IN}$ via ac coupling of C4-C0, such that $V_P = V_{COM}-V_{IN}$.

After the sampling is completed, it enters the conversion phase. In this phase, the SAR logic generates a control signal to switch S4 from ground to the reference voltage ($V_{REF}$) , which supplied by the band gap reference, to increase $V_P$ by $(8/16) \times V_{REF}$. Thus, $V_P = V_{COM}-V_{IN}+1/2V_{REF}$. $V_P$ is compared with $V_N$ to determine whether $V_P$ is higher or lower than $V_N$, which is essentially the comparison of $V_{IN}$ and 1/2$V_{REF}$. The output of the SA is sent to the SAR logic to determine the control of the capacitor switches in the next operation phase. When $V_P$\textgreater $V_N$, it means that $V_{IN}$\textless $1/2V_{REF}$, in the next operation phase, the SAR logic generates a control signal to switch S4 from $V_{REF}$ to ground and switch S3 from ground to $V_{REF}$ to decrease $VP$ by ${(4/16-8/16) \times V_{REF}}$. Thus, $V_P = V_{COM}-V_{IN}+1/4V_{REF}$. When $V_P$\textgreater $V_N$, the SAR logic generates a control signal to switch S3 from ground to $V_{REF}$ to increase $V_P$ by $(4/16) \times V_{REF}$. Thus, $V_P = V_{COM}-V_{IN}+1/2V_{REF}+1/4V_{REF}$. After 4 operation phases, the SAR logic will determine 4-bit digital code to get the final readout of the SAR-ADC.

In summary, analog in-MRAM computing was realized based on the above-mentioned proposal, using modified MRAM peripheral circuits implementation and device-circuit interaction design methods. Limited device TMR ratio can be addressed peripheral circuit within 1T-1M bit-cell based MRAM macro. The m-TMR formed by latching structure can meet the requirements of analog calculation. Circuit nonlinearity can be alleviated by the proposed CMF, which adds two additional transistors outside the current mirror. Multi-bit computing configuration allows analog computation by splitting data, converting it to a digital signal and then producing the result by a shift adder.

\begin{figure*}[!t]
    \centering{
    \includegraphics[width=1\textwidth]{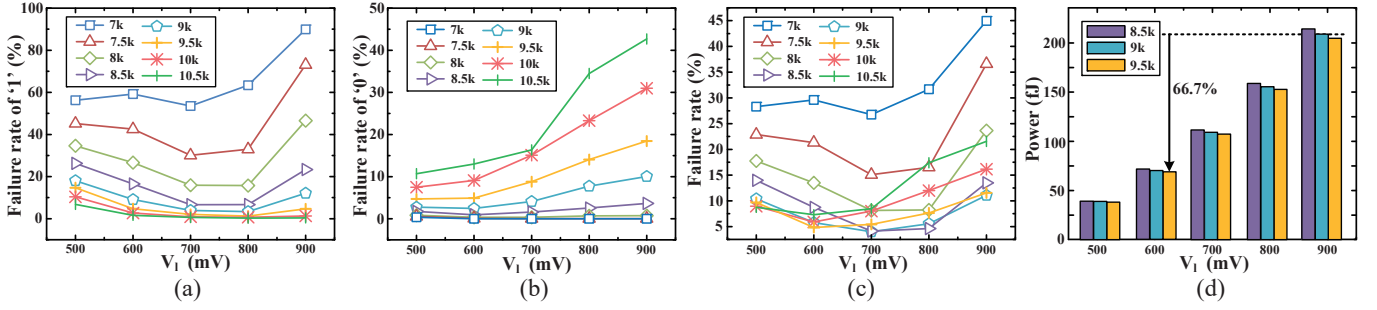}
    \caption{Monte-Carlo simulation results of different $V_{l}$ and reference resistance $R_{ref}$ (node W voltage greater than 750mV is `1'; Less than 150mV is `0'; Other voltages are latch failure): (a) Failure rate of latch data `1'. (b) Failure rate of latch data `0'. (c) Average failure rate of latch data. (d) Latch power consumption of 1-bit weight (holding time is 4ns).}
    \label{fig:7}}
\vskip -2pt
\end{figure*}

\section{Experimental Results}

The proposed m-TMR enlargement approach, analog in STT-MRAM computing circuits and system are verified and simulated with Spectre in Cadence Virtuoso front-end, using 28-nm CMOS process design kits and MTJ VerilogA compact model~\cite{model1,model2}. The simulation conditions are typical-typical (TT) corner, 27$^{\circ}$C, and the supply voltage ($V_{dd}$) is 900 mV. Table~\ref{Table3} lists fundamental MTJ parameters used in our analysis. MTJ physical parameter conforms to a commercial 40-nm critical dimension (CD) MTJ physical parameters. The effective low-resistance $R_{P}$ of MTJ is 6 k$\Omega$ with spin aligned ferromagnetic electrodes. An initial 200\% TMR ratio is configured.

\begin{table}[ht]
\centering
\caption{Physical parameters of STT-MTJ for in-MRAM computing performance simulation}
\setlength{\tabcolsep}{3pt}
\begin{tabular}{|p{45pt}<{\centering}|p{95pt}<{\centering}|p{80pt}<{\centering}|}
\hline
Parameters         & Description               & Default Value\\\hline\hline
$\Delta {H_0}$   & Activation energy      & 0.8eV           \\\hline
$\Gamma$         & Field acceleration      & 1.7cm/MV\\\hline
$\beta$              & Shape parameter      & 1.5          \\\hline
$k_B$             & Boltzmann constant   & 8.625 $\cdot$ $10^{ - 5}$eV/K\\\hline
${T_0}$            & Ambient temperature & 300K\\\hline
Variable             & Description               & Default Value\\\hline
$t_{ox}$       & Oxide barrier thickness of & 0.85nm\\\hline
TMR(0)             & TMR with 0 stress voltage & 200\%\\\hline
Area                 & MTJ surface                    & 40nm $\cdot$ 40nm $\cdot$ $\pi$/4\\\hline
${t_{sl}}$         & Thickness of free layer     & 1.3nm\\\hline
${V_{sl}}$        & Volume of free barrier     & Area $\cdot$ ${t_{sl}}$\\\hline
\end{tabular}
\label{Table3}
\end{table}

\subsection{Power consumption of latch structure}

The reference resistance $R_{ref}$ and the latching voltage ($V_{l}$) in latch structure affect the yield and power consumption. In order to reduce power consumption and improve latch variability performance, 5000 Monte-Carlo runs analysis is performed for $R_{ref}$ and $V_{l}$ under different TMR, and the 1-bit latch power consumption under different conditions is evaluated. Comprehensive analysis of the influence of $R_{ref}$ and $V_{l}$ on latching and selection of appropriate parameters can reduce power consumption and improve yield.

\begin{table}[ht]
\centering
\caption{Simulated latch performance with different TMR ratio}
\setlength{\tabcolsep}{3pt}
\begin{tabular}{|p{40pt}<{\centering}|p{40pt}<{\centering}|p{40pt}<{\centering}|p{40pt}<{\centering}|p{40pt}<{\centering}|}
\hline
TMR (\%)   & $\displaystyle \mathrm{R_{ref}}$ (k$\Omega$)  & $\displaystyle \mathrm{V_{l}}$ (mV)  & Power (fJ) & Yeild (\%)\\\hline
50   &   7.7   &   700  &   97.8  &   75.8  \\
100   &   8.5   &   600  &   78.4  &   86.8  \\
150   &   9   &   600  &   74.3  &   93.8  \\
200   &   9.5   &   600  &   70.8  &   95.2  \\
250   &   9.5   &   600  &  68.8  &   97.5  \\\hline
\end{tabular}
\label{Table4}
\end{table}

\begin{figure*}[!t]
    \centering{
    \includegraphics[width=1\textwidth]{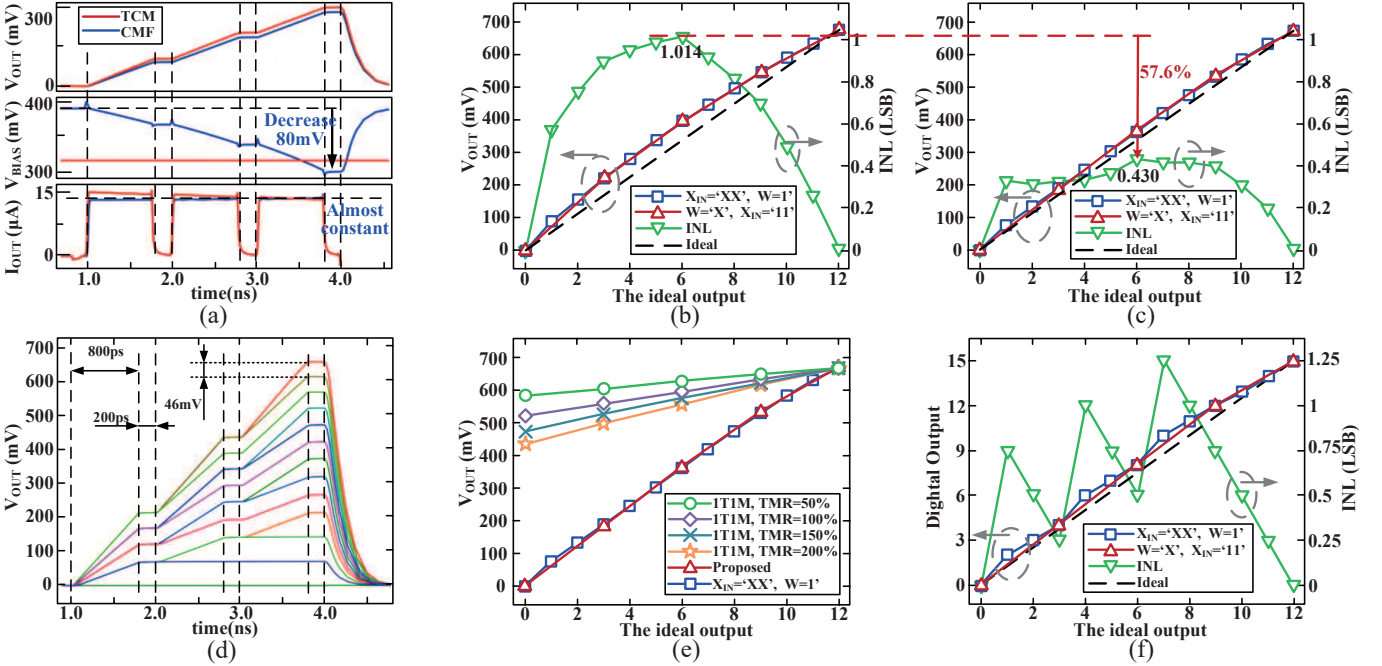}
    \caption{(a) Operation waveform of CMF and TCM block. (b) Output voltage of TCM (`X' is 1-bit any binary number, `XX' is 2-bit any binary number). (c) Output voltage of CMF. (d) Transient simulation results of output voltage (2bIN-1bW-4Acc, @TMR=200\%). The bottom to top waveforms correspond to the calculated results from 0 to 12. (e) Output voltage of the conventional 1T-1M array and the proposed 1T-1M array under different TMR. (f) Digital output quantized by the 4-bit SAR-ADC.}
    \label{fig:8}}
\vskip -2pt
\end{figure*}

Fig.~\ref{fig:7}(a) and \ref{fig:7}(b) show the failure probability of latch `1' and `0' under different conditions. In order to estimate the latching yield of in-MRAM computing, the voltage of the node $W$ greater than 750 mV is `1', less than 150 mV is `0', and all other voltages are latching faults. The yield is sensitive to the change of reference resistance $R_{ref}$, and the latch voltage $V_{latch}$ has different effects on the yield under different reference resistance $R_{ref}$. To ensure the overall yield of the latch structure, the two cases need to be considered comprehensively, and the average failure rate is shown in Fig.~\ref{fig:7}(c). When the reference resistance $R_{ref}$ is 8.5 k$\Omega$, 9 k$\Omega$ and 9.5 k$\Omega$, and the supply voltage $V_{l}$ is 700 mV, 700 mV and 600 mV respectively, the failure rate maintains a low level. Considering the power consumption simulation results in Fig.~\ref{fig:7}(d), the reduction of latch voltage $V_{latch}$ has a great impact on latch power consumption. Therefore, we select reference resistance 9.5 k$\Omega$ and supply voltage $V_{latch}$ to obtain 95.2\% accuracy and 70.8 fJ power consumption. Compared with the case of 900 mV power supply voltage as $V_{latch}$, the result of comprehensive analysis is that the failure rate is reduced by 56.4\% and the power consumption is reduced by 66.7\%.    

The optimal $R_{ref}$ and $V_{l}$ are obtained after trading off yield and power consumption, as summarized in Table~\ref{Table4}. The higher the TMR, the higher the yield and the lower the power consumption. After amplification by the inverter, the m-TMR is not related to the initial TMR ratio. TMR only affects the yield. Simulation results show that the m-TMR is about 15000. 

\subsection{Enhanced INL}

The contrast of waveforms between proposed CMF structure and traditional current mirror (TCM) is demonstrated in Fig.~\ref{fig:8}(a). $V_{BIAS}$ of the TCM stays the same, thus the charging current $I_{OUT}$ of the output capacitor decreases with the increase of output voltage $V_{OUT}$. As a comparison, $V_{BIAS}$ of the CMF decreases 80mV with with the increase of $V_{OUT}$, so $I_{OUT}$ can be kept constant to enhance the linearity of the results.

Fig.~\ref{fig:8}(b) and \ref{fig:8}(c) show the $V_{OUT}$ versus the ideal output of TCM and CMF. Integral nonlinearity (INL) refers to the difference between the transformation curve and the ideal transformation curve along the longitudinal axis, indicating the degree to which the actual curve deviates from the ideal curve. Comparing the INL of the TCM and the CMF, it can be found that adding feedback structure significantly improves output linearity. In Fig.~\ref{fig:8}(b), the maximum value of INL is 1.014 LSB; the maximum value of INL in Fig.~\ref{fig:8}(c) is only 0.430 LSB. The INL can be reduced by 57.6\% by using a feedback structure. 

The output voltage range is highly depended on the calculation margin. However, the nonlinearity of the analog calculation increases as the output voltage $V_{OUT}$ approaches the supply voltage $V_{dd}$ of 900 mV. The ability of the CMF structure to enhance linearity is limited. The maximum $V_{OUT}$ should be limited to 650 mV for regular CMF operation. The maximum $V_{out}$ can be adjusted by the voltage of the input pulse, $\gamma$ of the CMF structure and the capacitance of capacitor C1.

\subsection{Analog in-MRAM computing results}

The computation task with 4 accumulations of 2-bit inputs and 1-bit weights (2bIN-1bW-4Acc) is executed. The transition simulation results of $V_{OUT}$ is shown in Fig.~\ref{fig:8}(d). The waveform from bottom to top correspond to the calculated results 0 to 12 respectively. The charging time $T_H$ of each cycle is 800 ps. Around 4 ns, the difference between adjacent curves gradually decreases from bottom to top, and the minimum is 46 mV. This is due to the nonlinearity of the analog calculation. Before 4 ns, the SAR-ADC completes sampling of $V_{OUT}$. After 4 ns cycle, the SAR-ADC performs analog-to-digital conversion, the output capacitor discharges, and the array latches the next column of weight data.

In order to reflect the influence of proposed 1T-1M on the calculation, the in-MRAM computing structure of 1T-1M is also simulated under the same simulation conditions. Fig.~\ref{fig:8}(e) shows the $V_{OUT}$ of conventional 1T-1M array and proposed 1T-1M array under different TMR. For conventional 1T-1M array, as TMR increases, the influence curve of $W$ is close to that of $X_{IN}$. The difference between the two curves is too big to match together, and the analog calculation cannot be realized. On the contrary, the proposed 1T-1M uses latches to magnify the m-TMR by 7500 times when TMR equals 200\%, and the two curves almost coincide. Different TMR only affects the yield of latched results, but has no effect on the the influence curve of W. As shown in Fig.~\ref{fig:8}(f), digital output quantized by 4-bit SAR-ADC is close to the ideal output. The maximum INL is 1.25 LSB.

\subsection{Analysis of overall energy efficiency}

\begin{figure}[ht]
    \centering{
    \includegraphics[width=0.48\textwidth]{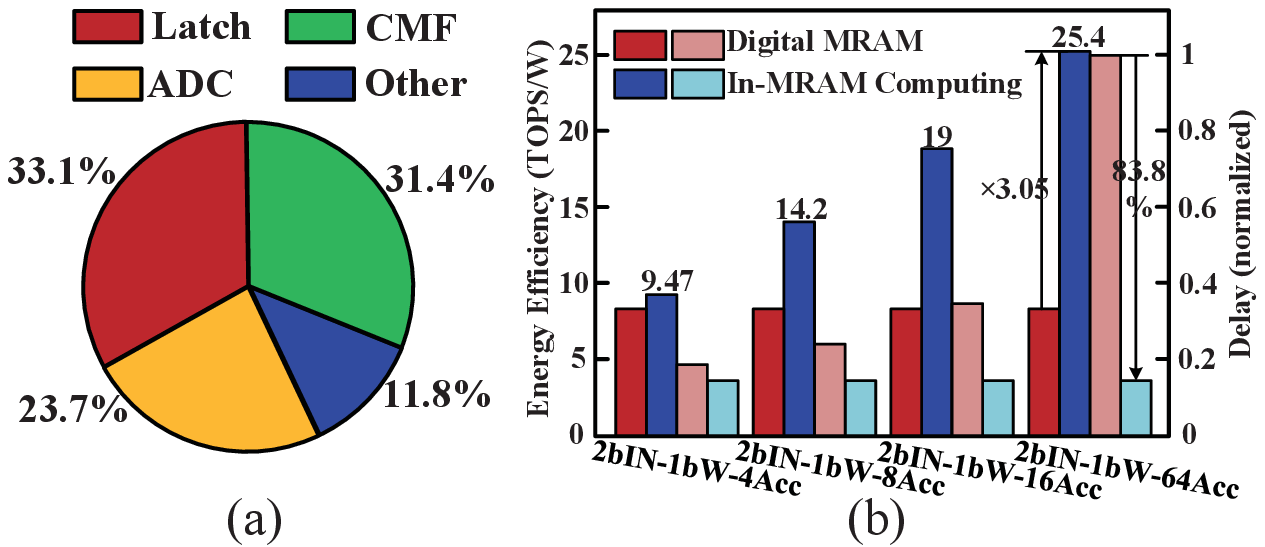}
    \caption{(a) Proportion of power consumption (2bIN-1bW-4Acc). (b) Energy efficiency and delay of in-MRAM computing versus conventional digital MRAM.}
    \label{fig:9}}
\vskip -2pt
\end{figure}

The power consumption of each building block is shown in Fig.~\ref{fig:9}(a). Latching power consumes the main part, more than 30\%, and the proportion of latching power increases as the number of rows on at the same time increases. The power consumption of the CMF is also more than 30\%. This is because SAR-ADC sampling needs to be completed when the CMF is charged, $V_{OUT}$ of the CMF structure need more driving capability. When the number of open lines increases, the current generated by each line is reduced by adjusting the voltage value of the IN pulse to keep the maximum $V_{OUT}$ constant. So the proportion of CMF structure power consumption does not increase as the number of rows on at the same time increases.

Fig.~\ref{fig:9}(b) shows the energy efficiency and in-MRAM computing latency versus conventional digital MRAM. As the number of open rows increases, the parallelism of in-MRAM computing increases, and the energy efficiency advantage becomes more obvious. When 64 rows open, the proposed structure achieves 25.4 TOPS/W which is 3.05 times more than conventional MRAM. The delay of in-MRAM computing cannot change as the amount of calculated data increases and is 83.8\% of digital MRAM when 64 lines are turned on. 

\section{Conclusion}

\begin{table}[ht]
\centering
\caption{Comparison of in-Memory Computing of Nonvolatile Memory}
\setlength{\tabcolsep}{3pt}
\begin{threeparttable}
\begin{tabular}{|m{37pt}<{\centering}|m{35pt}<{\centering}|m{35pt}<{\centering}|m{35pt}<{\centering}|m{35pt}<{\centering}|m{35pt}<{\centering}|}
\hline
 & ISCAS'21 \cite{flash}  & Nat. Com'20 \cite{PCM}  & ISSCC'20 \cite{RRAM1} & VLSI'20 \cite{mram-cim3} & This work\\\hline
Type &Flash   &   PCM   &   RRAM  &   SOT-MRAM  &   STT-MRAM  \\\hline
Process&180nm   & 90nm   &   130nm  &   22nm  & 28nm\\\hline
Bit-cell &1T1FGT   &   1PCM   &   2T2R  &  2T2M  &  1T-1M  \\\hline
Computing Bit & 4bIN-4bW-8bOUT &  8bIN-4bW-8bOUT  &  1bIN-TbW-1bOUT &  N/A  &  2bIN-1bW-4bOUT \\\hline
Energy Efficiency &37\tnote{1} TOPS/W   & 11.9 TOPS/W &   78.4 TOPS/W  &  N/A  &  9.47-25.4 TOPS/W  \\\hline
Write Voltage\tnote{2} & $>7$V  &   $>2$V   &  $>3$V  &   $<1$V  &   $<1$V  \\\hline
Process Deviation\tnote{2} & $\sigma<$15\%    &   $\sigma<$15\%   &  $\sigma>$15\%   &   $\sigma<$15\%   &   $\sigma<$15\%   \\\hline
\end{tabular}
 \begin{tablenotes}
        \footnotesize
        \item[1] Only MAC operation, other peripheral work was not included.
        \item[2] A part of the comparison refers to \cite{mram-cim3}
      \end{tablenotes}
 \end{threeparttable}
\label{Table5}
\end{table}
The performance of recent in-memory computing of NVMs is compared in Table~\ref{Table5}. Indeed, STT-MRAM based IMC shows little superiority of energy efficiency over the other NVMs. This is mainly due to its intrinsic low TMR ratio. The process of latching magnetoresistance difference to generate the amplified m-TMR requires at least one-third of the comprehensive analog in-MRAM computing power consumption. The advantages of STT-MRAM over other memory are low write voltage and low process deviation. Logic compatible writing voltage (e.g., 0.9 V with 28nm CMOS) can be applied for ultra-low power operation. Device variability induced unreliable effects can be partially alleviated when using STT-MRAM for analog IMC. 

A favorable scenario of in-MRAM computing could be artificial intelligence applications on Internet of things (IoT) edge devices, as normally-off and instant-on storage and computing units, e.g., emphasized energy cost in sleep-mode~\cite{ULPbench}. In this paper, an analog in-MRAM computing structure is proposed and experimentally realized. On the basis of commercial 1T-1M bit-cell, modified 1T-1M bit-cell builds the latch structure through the peripheral circuit. The m-TMR obtained from simulation can reach a maximum of 15000, which meets the requirements of analog calculation. The proposed CMF adds feedback current stabilization on the basis of traditional current mirror, and the INL is reduced by 57.6\%. The energy efficiency of the entire computing architecture reached 25.4TOPS/W (2bIN-1bW-64Acc), which is 3.05$\times$ that of the traditional digital MRAM.


\ifCLASSOPTIONcaptionsoff
  \newpage
\fi



\bibliography{IEEEabrv,bare_jrnl}
\end{document}